\documentclass[%
 reprint,
showpacs,
 amsmath,amssymb,
 aps,
pre,
]{revtex4-1}

\usepackage{graphicx}
\usepackage{dcolumn}
\usepackage{bm}

\begin{document}

\title[aaa]{The significance of trends in long-term correlated records}

\author{Araik Tamazian}
\affiliation{Institut f\"ur Theoretische Physik, Justus-Liebig-Universit\"at Giessen, 35392 Giessen, Germany}

\author{Josef Ludescher}
\affiliation{%
Institut f\"ur Theoretische Physik, Justus-Liebig-Universit\"at Giessen, D-35392 Giessen, Germany
}%

\author{Armin Bunde}
\affiliation{%
Institut f\"ur Theoretische Physik, Justus-Liebig-Universit\"at Giessen, D-35392 Giessen, Germany
}%

\begin{abstract}
  We study the distribution
$P(x;\alpha,L)$ of the relative  trend $x$ in long-term correlated records of
length $L$ that are  characterized by a Hurst-exponent $\alpha$ between 0.5 and 1.5 obtained by DFA2. The relative trend $x$ is the ratio between 
 the strength of the trend $\Delta$ in the record measured by linear regression,
and the standard deviation $\sigma$ around the regression line. We consider $L$ between 400 and 2200,
which is the typical length scale of monthly local and annual reconstructed global climate records.
Extending previous work by Lennartz and Bunde \cite{Lennartz2011} we show explicitely that $P$ follows the student-t distribution
$P\propto [1+(x/a)^2/l]^{-(l+1)/2}$, where the scaling parameter $a$ depends on both $L$ and $\alpha$, while the effective length $l$ depends,
for $\alpha$ below 1.15, only on the record length $L$. From $P$ we can derive an analytical expression for the trend significance
$S(x;\alpha, L)=\int_{-x}^x P(x';\alpha,L)dx'$ and the border lines of the $95\%$ percent significance interval. We show that the results are nearly independent of the distribution of the data in the record, holding for Gaussian data as well as for highly skewed non-Gaussian data.
For an application, we use our methodology to estimate the significance of Central West Antarctic warming.

\end{abstract}

\maketitle
\section{Introduction}
 The variability of a data set $\{y_i\}, i=1,\dots,L$, depends 
on its (internal) correlation properties and can be influenced by external mechanisms. Prominent examples are temperature records\cite{Lennartz2011, Mandelbrot, Bloomfield1992,  Pelletier1997, Koscielny1998, Malamud1999, Talkner2000, Weber2001,Monetti2003,Eichner2003, Fraedrich2003, Blender2003, GilAlana2005, Cohn2005, Kiraly2006, Rybski2006,Rybski2008, Zorita2008,Giese2007, Rybski2009,Halley2009, Lennartz2009a,Fatichi2009, Franzke2010, Franzke2012, Lovejoy2012, Franzke2013, Bunde2014}, river flows \cite{Hurst1951, Mandelbrot11968, Montanari2000, Montanari2003, Koutsoyiannis2003, Koutsoyiannis2006, Kantelhardt2006, Koscielny2006}, and sea level heights \cite{Beretta2005, Dangendorf2014, Lennartz2014} that all show a strong natural long-term persistence and, in addition, are effected by anthropogenic influences  that may lead to  an additional trend. 

In long-term persistent records,  small events have a tendency to cluster in "valleys" while large events tend to cluster in "mountains". Accordingly, long-term persistent records exhibit a pronounced valley-mountain structure, where it is difficult to distinguish a natural trend (starting in a valley and ending in a mountain) from a small  external deterministic trend. The problem of estimating the anthropogenic trend
in temperature records, river flows and sea level height  is an important issue in hydroclimatology and is called the "detection problem" \cite{Bloomfield1992, Hasselmann1993,Hegerl1996, Zwiers1999}.
The central quantity here is the probability $P(x;\alpha,L)$ that in a long-term persistent record of length $L$, characterized by the Hurst exponent $\alpha$, a relative trend of strength $x$ occurs;
$x$ is obtained from the standard linear regression analysis of the record and represents the height $\Delta$ of the regression line divided by the standard deviation of $y_i$ around the regression line (see Section II). From $P$ one obtains the significance of the trend as well as its error bars (denoting the 95\% significance interval)\cite{Bloomfield1992, Lennartz2011,Lennartz2009a}.

In many previous attempts to solve this problem  (see, e.g., \cite{Santer2000, Turner2005, Steig2009, Bromwich2012, Bromwich2014b, IPCC2013}) it had been  {\it tacitly assumed}  that the persistence of the climate records can be modelled by an auto-regressive process of first order (AR(1)) which allowed, in a simple and straightforward way, to estimate the significance of the warming trend $\Delta$ and its error bars.
In this case, the central quantity  is the probability $P(x; C(1),L)$ that in an AR(1) record of length $L$, characterized by the detrended lag-1 autocorrelation $C(1)$, a relative trend of strength $x$ occurs.  It is well known that $P(x; C(1),L)$ follows a student-t distribution (see Eq (6)), where the scaling paramer $a$ and the effective length $l$ are functions of $C(1)$ and $L$. This approach, however, which is conventional in climate science, can only be considered as a crude approximation since the temperature variability cannot be described by an AR(1) process where the autocorrelation function $C(s)$ decays exponentially with time lag $s$, but is described by a long-term persistent process where  $C(s)$ decays algebraically with $s$.

In recent years, there have been several attempts to solve the detection problem in long-term persistent data \cite{Cohn2005,Rybski2006,Giese2007,Zorita2008, Rybski2009, Halley2009,  Lennartz2009a, Lennartz2011}.  
 Using Monte Carlo simulations and scaling arguments it was found empirically \cite{Lennartz2009a,Lennartz2011} that for  long-term persistent Gaussian data,  $P(x;\alpha,L)$ can be approximated reasonably by a Gaussian for small $x$ and by a simple exponential for large $x$ values.

Here we perform the same kind of calculations as in \cite{Lennartz2009a,Lennartz2011}, but with a considerably better statistics, and show that the best approximation for $P$, in the {\it whole $x$-regime}, is again the {\it student-t distribution}, where now the scaling parameter $a$ and the effective length $l$ depend on $\alpha$ and $L$. 
While the previous result \cite{Lennartz2011} represents a good approximation in the respective $x$-windows, the present result is more satisfying since it shows 
that the distributions of a relative trend in uncorrelated, short-term correlated and long-term correlated data 
all follow the same equation, namely a student-t-distribution, but with different scaling parameters $a$ and effective lengths $l$. Accordingly, the exceeding probability $W$ and the significance $S$ of a trend is described, in all these different systems, by the same hypergeometric function.  In addition, we also study $P$ for strongly skewed non-Gaussian data and find that, to a very good approximation, $P$ is the same for all considered distributions.
Finally, we apply our methodology to the West-Antarctic temperature record at Byrd station.

The paper is organized as follows: In Section II we describe how the  exceedance probability $W$ and the significance $S$ is related to $P$ and which form 
these quantities have for uncorrelated Gaussian noise and short-term correlated Gaussian data characterized by an AR(1) process. We also give a brief introduction into long-term persistent data and their characterization. In Sections III we present our numerical results for the significance of a relative trend in long-term persistent Gaussian (Section III) and non-Gaussian data (Section IV). In Section V we show how our approach can be applied to monthly temperature records. As an example we take the Byrd record from West Antarctica that has very recently been reconstructed \cite{Bromwich2014b}. In Section VI, finally, we summarize our results. 

\section{Detection of external trends}
 We consider a record $\{y_i\}, i=1,\dots,L$ and assume, without loss of generality, that the mean value $\bar y$ of the data is zero.
To estimate the increase or decrease of the data values in the considered time window of length $L$, one usually performes a regression analysis. From the regression line $r_i=bi + d$,  one obtains the magnitude of
the trend $\Delta y=c(L-1)$ as well as the fluctuations around the trend, characterized by the  standard deviation
$\sigma=[(1/L)\sum_{i=1}^L (y_i-r_i)^2]^{1/2}$. The relevant quantity we are interested in is the {\it relative trend}
\begin{equation}
x=\Delta y/\sigma. 
\label{x}
\end{equation}
When a certain relative trend has been measured in a data set, the central question is, if this trend may be due to the natural variability of the data set or not ("detection problem").  To solve this problem, one needs to know the probability $P(x;L)dx$ that in model records with the same persistence properties as the considered data set, 
a relative trend between $x$ and $x+dx$ occurs. The probability density function $P(x;L)$ is symmetric in $x$. In the following we consider $x>0$. From  $P$ we derive the {\it exceedance
 probability} $W(x;L)=\int_x^\infty P(x';L)dx'$
and the trend significance
\begin{equation}
S(x;L)=\int_{-x}^x P(x';L)dx'=1-2W(x;L).
\label{S}
\end{equation}
By definition, $S$ is the  probability that the relative trend in the record is between $-x$ and $x$.

If the significance of a relative trend is above 0.95  (or $95\%$), one usually assumes that  the considered trend cannot be fully explained by the natural variability of the record. 
The relation $S(x_{95}; L) = 0.95$ defines the upper and lower limits $\pm\,x_{95}$ of the $95\%$ significance interval (also called confidence interval). By the above
assumption, relative trends $x$ between $-x_{95}(L)$ and $x_{95}(L)$ can be regarded as natural. 
If $x$ is above $x_{95}$, the part $x-x_{95}$ cannot be explained by the natural variability of the record and thus can be regarded as minimum external relative trend,
\begin{equation}
x_{\rm ext}^{\rm min}=x-x_{95}.
\label{x_min}
\end{equation}
On the other hand, the external trend cannot exceed
\begin{equation}
x_{\rm ext}^{\rm max}=x+x_{95},
\label{x_max}
\end{equation}
which thus represents the maximum external relative trend.
By definition, $x_{\rm ext}^{\rm min}$ represents the lower margin of the observed relative trend 
that cannot be explained by the natural variability alone, while $x_{\rm ext}^{\rm max}$
ist the largest possible external relative trend consistent with the natural variability of the record.
According to Eqs. (\ref{x_min}) and (\ref{x_max}), $\pm\, x_{95}(L)$ can be regarded as error bars for an external
relative trend in a record of length $L$.

\subsection{White noise}
For uncorrelated Gaussian data (white noise), it has been assumed (see \cite{Santer2000} and references therein) that the ratio $t_b$ between
the estimated trend slope $b$ and its standard error $s_b$ (see Eqs. (1-5) in \cite{Santer2000}) follows a student-t distribution.
This assumption can be written as
\begin{equation}P(x;L)= \frac{\Gamma(\frac{l(L)+1}{2})}{\Gamma(\frac{l(L)}{2})\sqrt{\pi l(L)}a} \left(1+\frac{(x/a)^2}{l(L)}\right)^{-\frac{l(L)+1}{2}},
\label{student}
\end{equation}
with the degrees of freedom
 \begin{equation}l(L)=L-2\label{l}\end{equation}
 and the scaling parameter 
 \begin{equation}a=\frac{\sqrt{12}(L-1)}{\sqrt{L^2+ 2}} \frac {1}{\sqrt{l(L)}}\label{a_0}.\end{equation}
 $\Gamma$ denotes the $\Gamma$-function. In the limit of large $L$, $a$  tends to $a\cong\sqrt{12}/\sqrt{l(L)}$.

From (\ref{student}) and (\ref{S}) one can obtain straightforwardly  the  significance trend $S$ as a function of $x/a$ and $l(L)$,
\begin{equation}
\begin{split}
S(x;L)= 2\frac{x}{a}\frac{\Gamma(\frac{1}{2}(l(L)+1)}{\sqrt{\pi l}\Gamma(\frac{l(L)}{2})} \times \\
 {}_2 F_1 \left(\frac{1}{2},\frac{1}{2}(l(L)+1);\frac{3}{2};-\frac{(x/a)^2}{l(L)}\right) 
\end{split}
\label{student-S}\end{equation}
where $_2F_1$ is the hypergeometric function.

\subsection{Short-term correlations}
The most basic model for short-term correlations in data sets is the autoregressive process of first order (AR1), where the data satisfy the equation
\begin{equation}y_{i+1}=c_1y_i + \eta_i, i=1,2,\dots, L-1.\end{equation}
 Here, the AR1
parameter $c_1$ is between -1 and 1 and $\eta_i$ is white noise. For
$c_1<0$ the data are antipersistent, while for $c_1>0$ they are
persistent.  For $c_1=0$, they are white noise. 

 For characterizing the persistence of a record, one often studies the autocorrelation 
function  $C(s) = \langle y_i y_{i + s} \rangle \equiv {{1}\over{(L - s)}} \sum^{L - s}_{i = 1} y_i y_{i + s}/({1\over{L}} \sum^{L}_{i = 1} y_i ^2)$. 
By definition, $C(0)=1$. It is easy to show that for AR1 processes, in the limit of $L\to\infty$,
$C(s)$ decays exponentially, $C(s)=c_1^s$, i.e., $c_1$
is identical to the lag-1 autocorrelation $C(1)$. For $c_1>0$, $C(s)$ can be written as $C(s)=\exp(-s/s_x))$ where
$s_x=1/\vert\ln c_1\vert$ denotes the persistence time.

It has been shown \cite{Santer2000} that for sufficiently large $L$ where $C(1)=c_1$, $P$  has approximately the form of the student-t distribution 
Eq.  (\ref{student}), 
with 
\begin{equation}l(L)= L\  \frac{ 1-C(1)}{1+C(1)}-2 \label{l-1}\end{equation}
and 
 \begin{equation}a=\frac{\sqrt{12}(L-1)}{\sqrt{L^2+ 2}} \frac {1}{\sqrt{l(L)}}\cong\frac{\sqrt{12}}{\sqrt{l(L)}}\label{a-1}\end{equation}
 Accordingly, the significance $S$ of the trend is described by Eq. (\ref{student-S})  with $l(L)$ from  (\ref{l-1}) and $a$ from  (\ref{a-1}).

\subsection{Long-term persistence}
Long-term correlated records can be characterized by the power spectral density $S(f)=\vert y(f)\vert^2$, where $\{y(f)\}$, $f=0,1,\dots ,L/2$, is the Fourier transform of $\{y_i \}$. With increasing frequency $f$, $S(f)$ decays by a power law,
\begin{equation}
S(f)\sim f^{-\beta},
\label{S(f)}
\end{equation}
where $\beta>0$ characterizes the long-term memory \cite{Pelletier1997}. For white noise, $\beta=0$. Records with $0<\beta<1$ can be  characterized
by an  autocorrelation function $C(s)$ that decays by a
power law, $C(s)\sim (1-\gamma)s^{-\gamma}$, with $\gamma=1-\beta$. To generate long-term persistent data, one usually uses the Fourier-filtering technique based on (\ref{S(f)}), 
where long records of 
uncorrelated Gaussian data (typically of length $L_0=2^{21}$) are transformed to Fourier space. The result is multiplied by $f^{-\beta/2}$ and then transformed back to time space. The resulting record is Gaussian distributed.
For obtaining records of the desired length $L$, one divides the long record into segments of length $L$.

Since both $S(f)$ and $C(s)$
exhibit large finite size effects and are strongly influenced by external deterministic trends, one usually does not use these methods to characterize the long-term persistence, but prefers methods like the detrending fluctuation analysis of 2nd order (DFA2) \cite{Kantelhardt2001} where linear trends in the data are eliminated systematically.
 
In DFA2 one measures the variability of a record by studying the fluctuations in segments of the record as a function of the length $s$ of the segments. Accordingly, one first divides the record  $\lbrace{y_i}\rbrace, i = 1, 2, \dots, L$,
into non-overlapping windows  $\nu$ of lengths $s$. 
 Then one focuses, in each segment $\nu$, on the cumulated sum ${Y_i}$ of the data and 
determines the variance $F^2_\nu (s)$ of the ${Y_i}$ around the best polynomial fit of order 2. 
 After averaging $F^2_\nu (s)$ over all segments $\nu$ and taking the square root, we arrive at the desired fluctuation function $F(s)$. 
One can show that 
\begin{equation}F(s)\sim s^{\alpha}.\end{equation}
The exponent $\alpha$ can be associated with the Hurst exponent, and is related to the correlation exponent $\gamma$ and the 
spectral exponent $\beta$ by $\alpha=(1+\beta)/2$ and $\alpha=1-\gamma/2$. For uncorrelated data, $\alpha=1/2$. The DFA2 technique gives reliable results for time scales $s$ between 10 and $L/4$ \cite{Kantelhardt2001}.

When applying DFA2 to short-term persistent data, the fluctuation function $F(s)$ approaches a power law, with  $\alpha=1/2$, for $s$ well above the persistence time $s_x$. For $s$ well below $s_x$ (which can only be the case for $b$ very close to 1), $\alpha$ is close to 1.5. 
The difference in the functional form of $F(s)$ allows to distinguish between short-term and long-term persistent processes.

Recently, it has been shown \cite{Lennartz2011} by Monte Carlo simulations that in long-term persistent data of length $L$, where the Hurst exponent $\alpha$ is determined by DFA2, the probability density $P(x;\alpha, L)$ of the relative trend $x$
can be reasonably approximated by a Gaussian for small $x$ and by a simple exponential for large $x$. Using scaling theory, an analytic expression for $W(x;L)$ has been obtained, as a function of $\alpha$, in the two $x$-regimes. 
Here we follow the same route as in \cite{Lennartz2011}, but with a better statistics, and
find that the best approximation for $P$, in the whole $x$ regime,  is the  student-t-distribution Eq. (\ref{student}), where the scaling parameter $a$ and the effective length $l$ depend on both $\alpha$ and $L$.  

 Accordingly,  the significance of a relative trend in long-term persistent records is described by the same hypergeometric function as for white noise and AR1 noise, only the parameters $l$ and $a$ are different.
We also show that
the results derived for Gaussian data hold, in an excellent approximation, also for data with a symmetric exponential distribution $D(y)=(1/2)exp(-\vert y\vert)$
as well as for strongly skewed distributions like the one-sided exponential and one-sided power-law distribution.

\section{Significance  of trends in long-term correlated Gaussian data sets}
For determining $P(x; \alpha, L)$  numerically, we follow \cite{Lennartz2011}. We use the Fourier-filtering technique \cite{Mandelbrot} to generate 800 synthetic records of length $2^{21}$, for 241 global Hurst exponents $\alpha^*$ ranging from $\alpha^*= 0.1$ to $\alpha^*=2.5$. We are interested in data sets with lengths $L$ between $400$ and $2200$ which correspond, in monthly temperature data sets, to data lengths between 33 and 183 years. Accordingly, we divided each data set into subsequences of lengths $L=400, 500, 600,\dots,2200$.  In each subrecord of length $L$, we used linear regression to determine (i) the local DFA2 Hurst exponent $\alpha$  as the slope of the regression line in a double logarithmic presentation of the fluctuation function $F(s)$  between $s=10$ and $L/4$ and (ii) the relative trend $x$.  We are interested in $\alpha$ values between 0.5 and 1.5, which are most common in nature.

It has been noticed before \cite{Rybski2008, Lennartz2009a, Lennartz2011}, that the local Hurst exponents $\alpha$ obtained in each subrecord are not identical to the global  Hurst exponent $\alpha^*$ of the entire record, but vary around $\alpha^*$. The distribution of the local Hurst exponents $\alpha$,  for fixed $\alpha^*=0.75$
and  $L=400$ and 2200, is shown in Fig. 1a. As expected, the distribution narrows with increasing subrecord length $L$. Accordingly, when in a subrecord a certain local Hurst exponent $\alpha$ is measured, the subrecord may  be part of a long data set with a different global Hurst exponent $\alpha^*$.
Figure 1b shows, for fixed $\alpha=0.75$, the distribution of the $\alpha^*$  values. Again, the distribution narrows with increasing $L$.  
\begin{figure}[]
\begin{center}
\includegraphics[width=8.3cm]{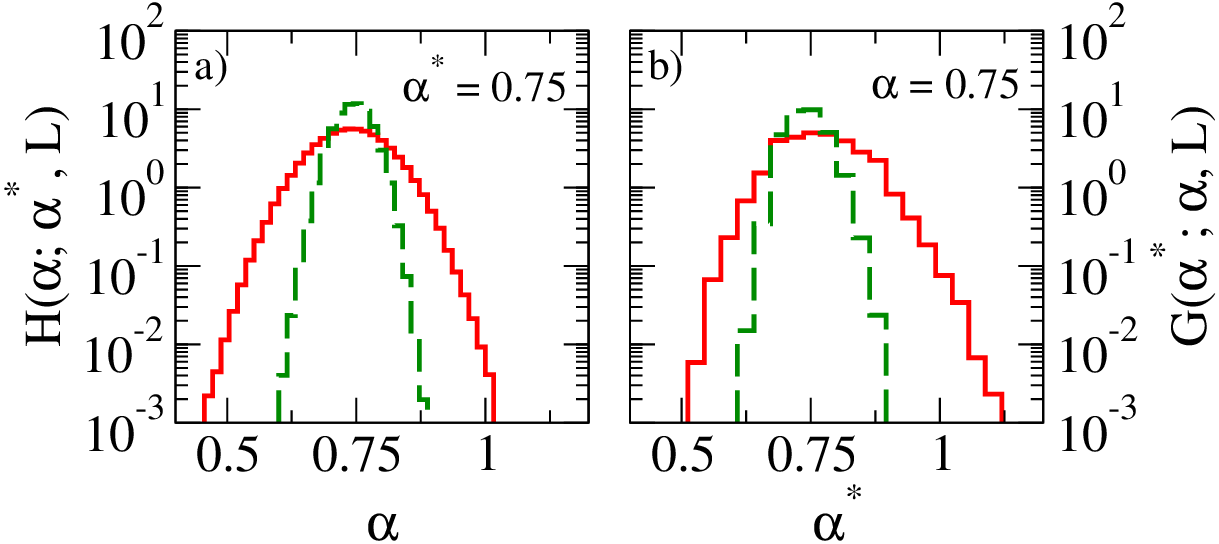}
\caption{(color online) 
(a) When long data sets with global Hurst
exponents $\alpha^*$ (created by Fourier filtering) are divided into sub-segments of length $L$, the local Hurst exponents $\alpha$ vary around $\alpha^*$. The figure shows the distribution  $H(\alpha; \alpha^*,L)$ of the local Hurst exponents $\alpha$ (obtained by DFA2) in segments of lenght $L=400$ (continuous red line) and $L=2200$ (dashed green line), for fixed 
global Hurst exponent $\alpha^*=0.75$. 
(b) According to (a), the same local Hurst exponent $\alpha$ can originate from  long  data sets with different  global Hurst
exponents $\alpha^*$. The figure shows, for the local Hurst exponent $\alpha=0.75$ in segments of length $L=400$ (continuous red line) and $L=2200$ (dashed green line),  the distribution  $G(\alpha^*; \alpha,L)$ of the involved global Hurst exponents $\alpha^*$.
}
\label{1}
\end{center}
\end{figure}
As a consequence, for determining the significance of a relative trend in a  long-term correlated record of length $L$, one cannot simply
identify the local Hurst exponents $\alpha$ with the global one, but has to determine the local Hurst exponents in each subrecord separately. As we will show in the second part of this Section, ignoring this fact will lead to a strongly enhanced significance. We like to note that a similar problems occurs in short-term persistent records, where only in long data sets the lag-1 autocorrelation function $C(1)$ is equal to the persistence parameter $b$. In  subrecords (or short data sets) of length $L$, the values of $C(1)$ fluctuate around $b$, and Eqs. (\ref{l-1}) and (\ref{a-1}) are not valid \cite{Bunde2014}.

 After having obtained in each subrecord $k$ (of fixed length $L$) the local Hurst exponents $\alpha_k$, we focus on those subrecords that have local $\alpha$ values  between 0.49 and 1.51. We divide the local $\alpha$ values into  51 windows of length 0.02, such that in the first window  $\alpha=0.5\pm0.01$, in the second  window  $\alpha=0.52\pm0.01$, and in the last  window $\alpha=1.5\pm0.01$. Then we determine,
in each  $\alpha$-window, the distribution $P(x; \alpha, L)$ of the relative trends as well as  the trend significance
$S$. 

\begin{figure}[]
\begin{center}
\includegraphics[width=8.3cm]{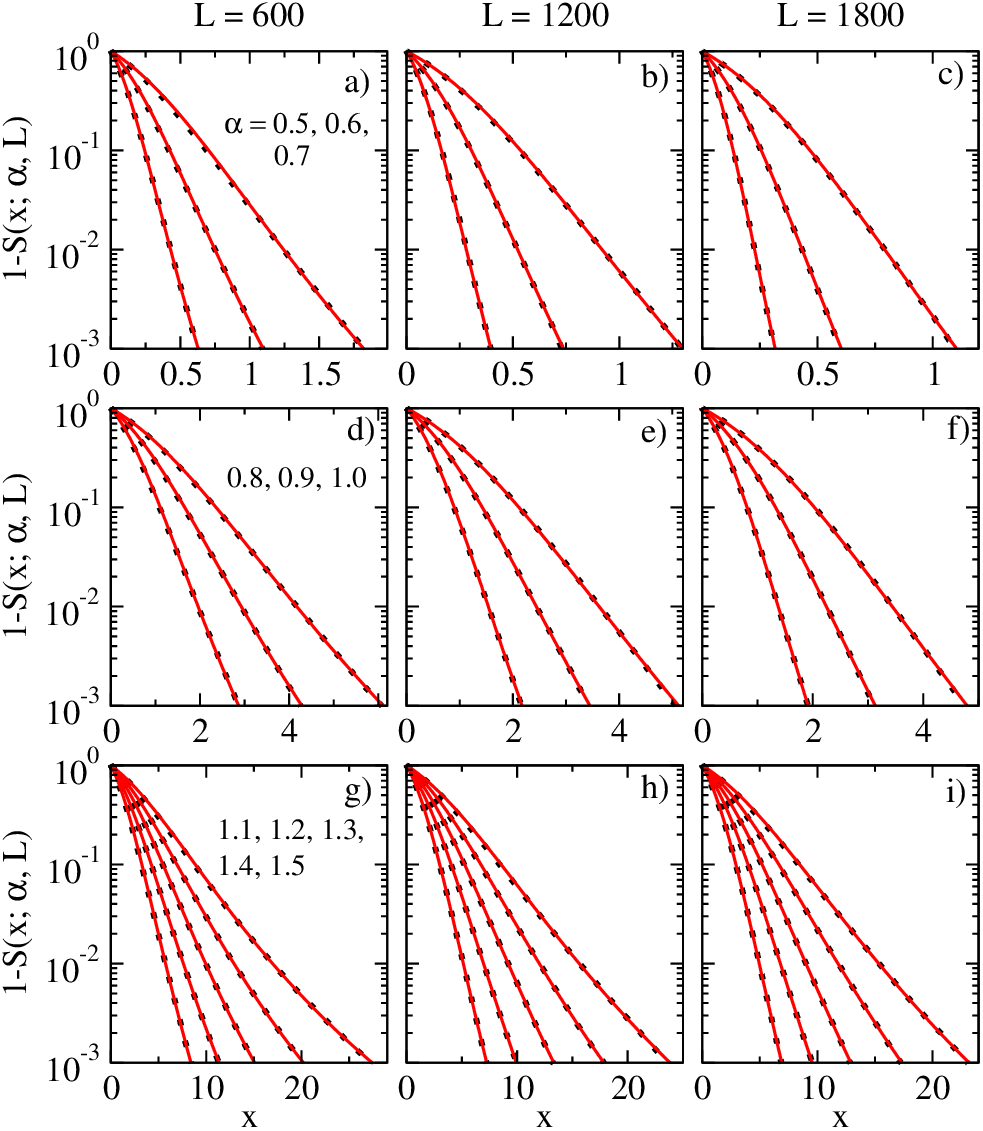}
\caption{(color online) Significance $S(x;\alpha,L)$ of relative trends $x$ occurring  
in long term correlated data of length $L$ and Hurst exponent $\alpha$. The data are Gaussian distributed. For clarity, we focus on $1-S$. (a) is  for $L=600$ and $\alpha=0.5, 0.6., 0.7$  (symbols from bottom to top). The continous lines show the corresponding student-t-fits. (b,c) Same as (a) but for the Hurst exponents $\alpha=0.8,0.9,1.0$ and $\alpha=1.1, 1.2, 1.3, 1.4, 1.5$, 
respectively. (d-f) and (g-i): Same as (a-c) but for record lenghts $L=1200$  and $L=1800$, respectively.
}
\label{2}
\end{center}
\end{figure}
Figure 2 shows  $1-S(x;\alpha,L)=2W(x;\alpha,L)$  for three representative data lengths $L=600$, 1200, and 1800, which in monthly climate records correspond to 50, 100, and 150 years.  The dots are our numerical results. The continous lines result from a fit of $S$
to Eq. (\ref{student-S}), with appropriately chosen values for the scaling parameter $a$ and the effective length $l$. The figure shows that over all three decades of $1-S$ considered here (where $S$ ranges from 0 to 0.999 (or from 0 to 99.9 \%)), the fit is excellent. The  parameters $l$ and $a$ are listed, for $L$ between  400 and 2200, in the Appendix. 

  Table 1 in the Appendix shows that for fixed record length $L$, the effective length $l$ is a constant in the most relevant range between $\alpha=0.5$ and 1.1 For example, for $L=600$, 1200, and 1800, $l =$ 9.24, 12.05, and 13.69, respectively. For $\alpha$ above 1.1, $l(L)$ decreases strongly.

  Figure 3 shows that the scaling parameter $a$ listed in Table 2 in the Appendix, can be approximated, for $\alpha$ between 0.5 and 1.2, by a power law, where the slope increases with increasing $L$. Above $\alpha=1.2$, $a$ shows only a very weak $L$ dependence.

\begin{figure}[]
\begin{center}
\includegraphics[width=6.3cm]{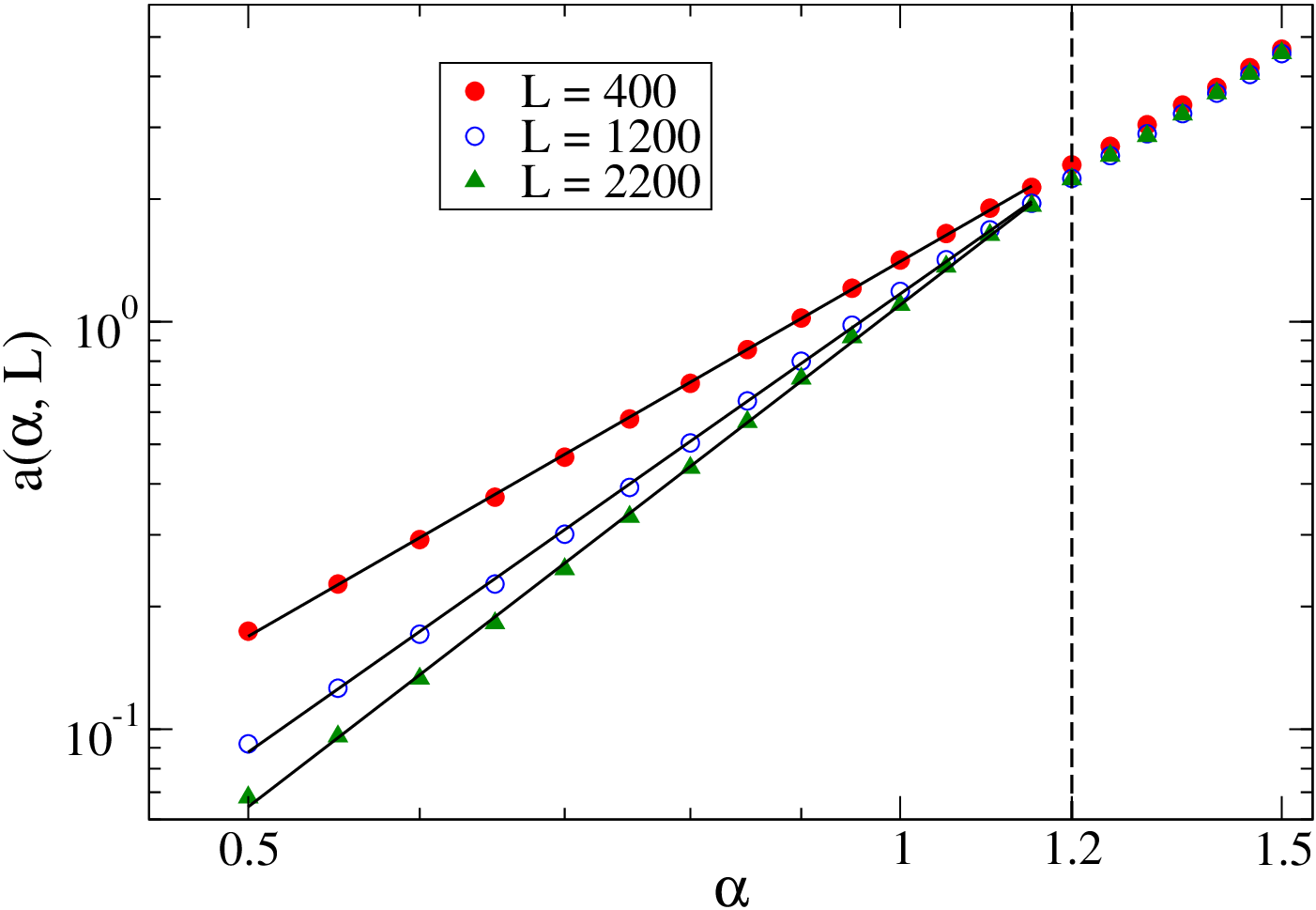}
\caption{(color online) The characteristic relative trend $a(\alpha,L)$ that characterizes the decay of $1-S$ (see Fig. 2 and Eq. (\ref{student-W}), as a function of the Hurst exponent $\alpha$ for three record lengths $L=400, 1200, 2200$. The straight lines are power law fits between $\alpha=0.5$ and 1.15.
}
\label{3}
\end{center}
\end{figure}
From Fig. 2, by intersecting $1-S$ with the constant $5\times10^{-2}$, we  obtain immediately  the relative trend $x_{95}$ that is conventionally used to estimate the error bars of a measured relative trend. Figure 4 shows $x_{95}$ for  $L=600$, 1200 and 1800. From the figure, one can immediately read off the error bars of a relative temperature trend in monthly records of length 50, 100, and 150y with DFA2 exponent $\alpha$, and specify its lower and upper bounds $x_{\rm ext}^{\rm min}=x-x_{95}$ and $x_{\rm ext}^{\rm max}=x+x_{95}$.
\begin{figure}[]
\begin{center}
\includegraphics[width=6.3cm]{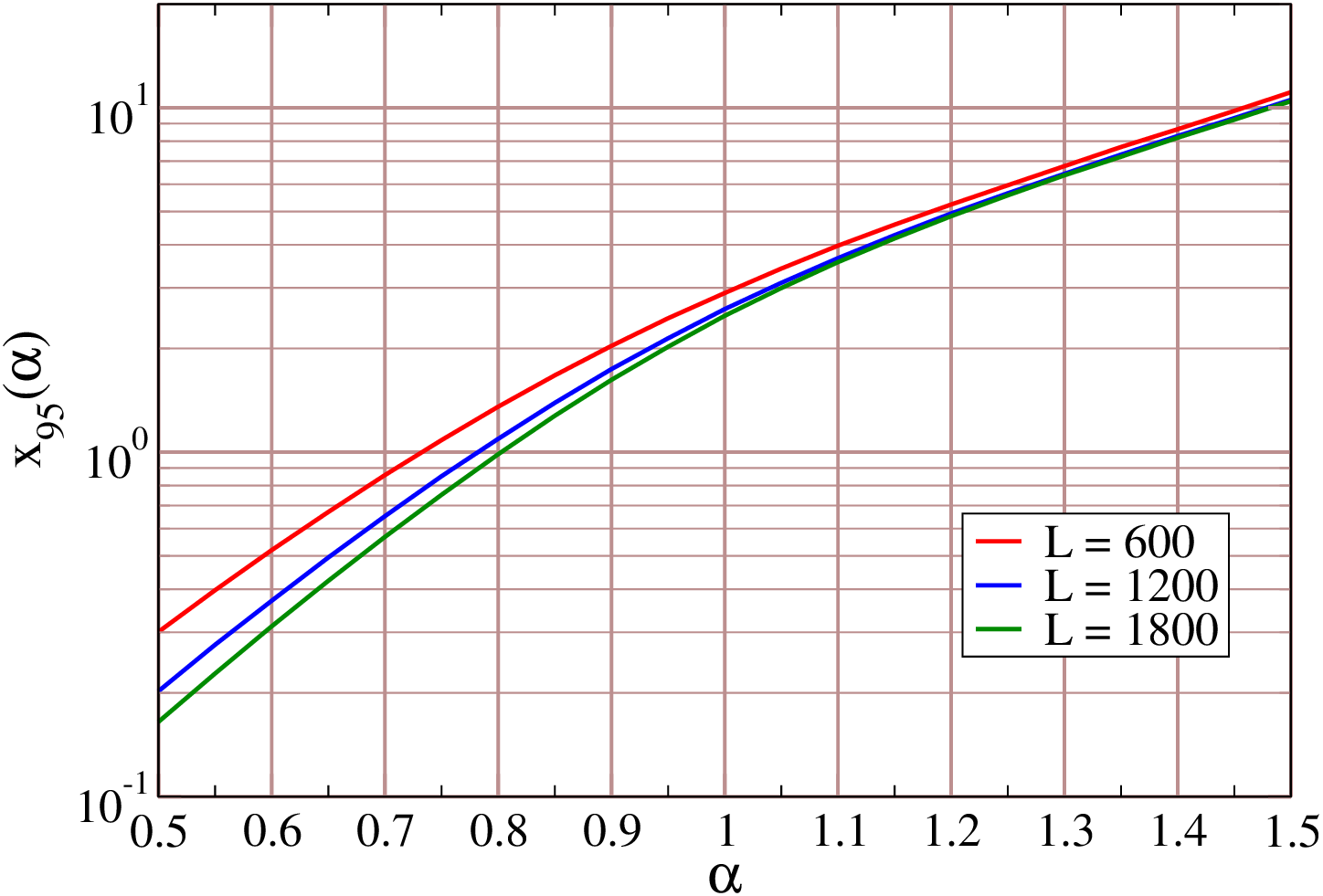}
\caption{(color online) The relative trend $x_{95}$ corresponding to the $95\%$ significance level as a function of the Hurst exponent $\alpha$ for the segment lengths $L=600, 1200$, and $1800$ (from top to bottom). By definition, $x_{95}$ is obtained from the intersection of $1-S$ in Fig. 2 with 1-0.95=0.05.
}
\label{4}
\end{center}
\end{figure}

Finally, at the end of this Section, let us go back to its beginning and ask the following question: Given a long record of length $L_0$ described by the global Hurst exponent $\alpha^*$, which is divided into subrecords of length $L\ll L_0$. What is the significance $\tilde S(x;\alpha^*,L)$ of a relative trend $x$ in these
subrecords? We expect that for large $L$ where all local Hurst exponents $\alpha$ are very close to $\alpha^*$,  $\tilde S(x,\alpha^*,L)$ and $S(x,\alpha,L)$ will coincide. For small $L$ we expect that $\tilde S(x,\alpha^*,L)$ overestimates the significance.

 The difference between $S(x,\alpha,L)$ and $\tilde S(x,\alpha^*,L)$ may also be regarded as follows: If we know a priori that the considered data set is characterized by a certain Hurst exponent (which is the case, for example, when we consider records consisting of Gaussian distributed independent random numbers or their cumulated sum), then $\tilde S(x,\alpha^*,L)$
gives the proper significance. 
  If we do not know the characteristics of the data set a priori,  the uncertainty is increased and we have to determine explicitely its Hurst exponent, 
  $S(x,\alpha,L)$ gives the proper significance.

Figure 5 shows $\tilde S(x,\alpha^*,L)$ for the global Hurst exponents $\alpha^*=0.5$, 0.75, 1, and 1.25 in subrecords of lengths $L=400, 1200,$ and 2200. The figure shows also the significance $S(x,\alpha,L)$, for  $\alpha=0.5$, 0.75, 1, and 1.25.  For $\alpha^*=0.5$, $\tilde S(x,\alpha^*,L)$ follows Eq. (\ref{student-S}) with (\ref{l}) and (\ref{a_0}). We found that also for $\alpha=0.75$ and 1, $\tilde S(x,\alpha^*,L)$ is well described by the student-t distribution (\ref{student-S}) with 
$l(L)=L-2$ (\ref{l}). For $\alpha=0.75$, the $a$ values are  0.512, 0.383, and 0.328 for $L=400$, 1200 and 2200, respectively. For $\alpha=1$, the respective  $a$ values are  1.313, 1.185, and 1.135.  
As expected, $\tilde S(x,\alpha^*,L)$  overestimates the significance of an observed relative trend and thus underestimates  the error bars $\pm x_{95}$ of a relative trend. For example, when in a monthly temperature record of length 600 (corresponding to 50 years) characterized by $\alpha=0.75$ a relative trend $x=1$ is measured, the proper significance of this trend is $S=0.94$, i.e. the trend is not significant. However, if falsely $\tilde S$ is used for estimating the significance, one overestimates the significance, since $\tilde S=0.975$ in this case.

\begin{figure}[]
\begin{center}
\includegraphics[width=8.3cm]{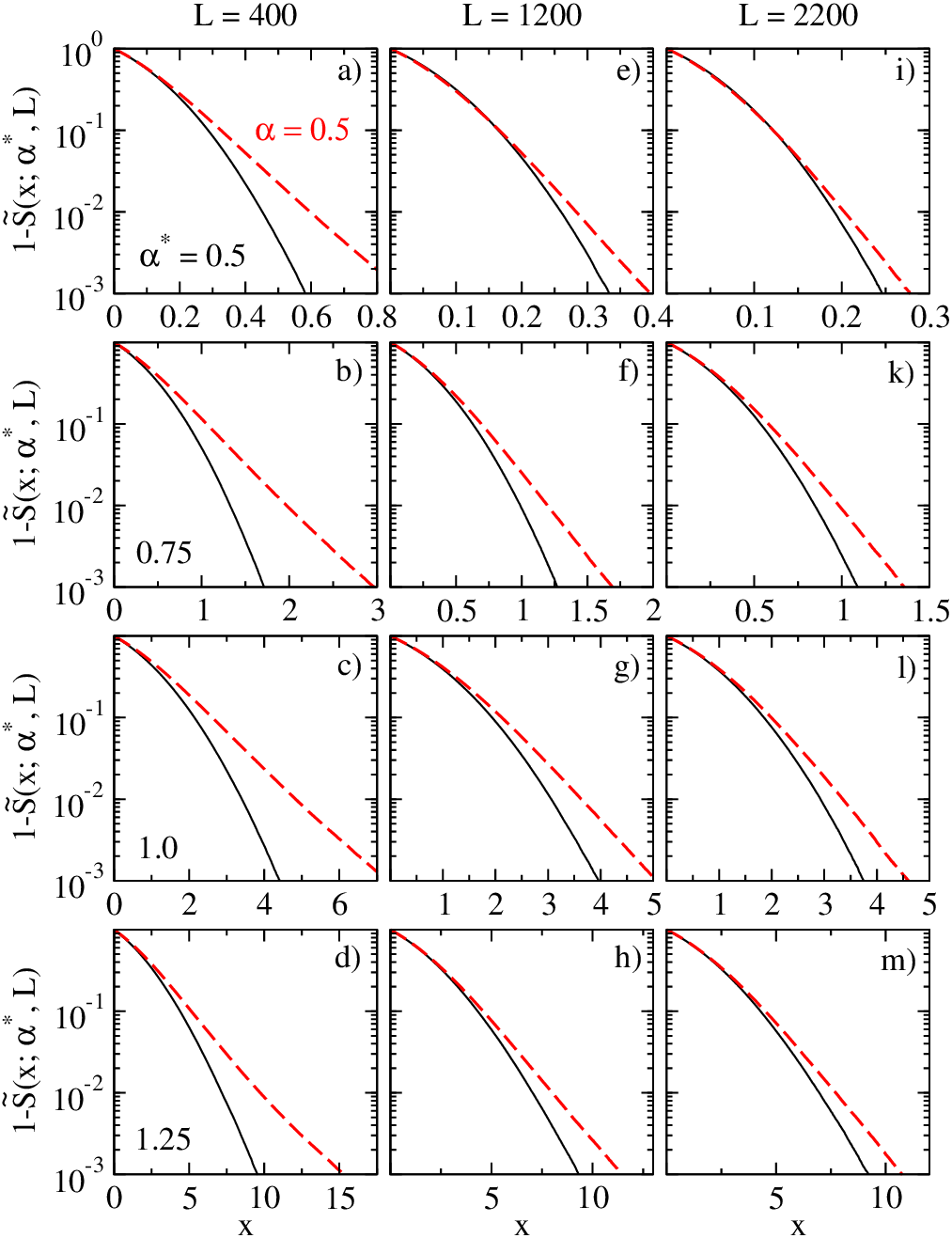}
\caption{(color online) Significance $\tilde S(x,\alpha^*,L)$ of relative trends $x$ occurring  
in segments of length $N$ of long data sets (length $2^{21}$) with fixed global Hurst exponent $\alpha^*$  (continous line). The data follow Gaussian  distributions. For clarity, we focus on $1-S$, as in Fig. 2.
We compare $\tilde S(x,\alpha^*,L)$ with the significance $ S(x,\alpha,L)$ of relative trends $x$ occuring in data sets of the same length $N$ with fixed local Hurst exponent $\alpha$ (dashed red line) that was shown in Fig. 2.
The considered segment lengths are $L=400, 1200,$ and 2200, shown in columns $1-3$. The Hurst exponents $\alpha^*=\alpha$ are 0.5, 0.75, 1.0, and $1.25$, shown in rows $1-4$. 
}
\label{5}
\end{center}
\end{figure}

\section{Significance of trends in long-term correlated non-Gaussian data sets}
By using the Fourier-filtering technique we generated long-term correlated Gaussian data $\{ y_i\}$.
Many natural records, e.g., monthly temperature anomalies where the seasonal trend has been removed, are Gaussian distributed.
But others, like river run-off data, have a quite skewed distribution and cannot be characterized by a Gaussian \cite{Kantelhardt2006, Koscielny2006, Bunde2013}. Accordingly, the question arises,
to which extend our results derived in the previous subsection are general and apply also to non-Gaussian distributions.

To answer this question, we have considered three kinds of non-Gaussian distributions: (i) the symmetric exponential distribution
$D(y)=(1/2) \exp(-\vert y \vert)$, (ii) the (highly skewed) exponential distribution $D(y)=\exp(-y), y\ge 0$, and (iii) the (highly skewed) power law distribution $D(y)=y^{-5}, y\ge 0$. To generate these distributions, we have first generated  long-term correlated data $\{ y_i\}$ of length $2^{21}$ that are Gaussian distributed, as above. Then we generated $2^{21}$ data of the considered non-Gaussian distribution and exchanged the long-term correlated Gaussian data rankwise
by the non-Gaussian data.

 By this simple  exchange technique we obtain long-term correlated data following the considered non-Gaussian distribution, but the global Hurst exponent as well as the local ones usually differ slightly from the original one. These slight deviations do not play a role here, since 
we consider $\alpha^*$ values between 0.1 and 1.9 and 
only the {\it local} $\alpha$ values  measured by DFA2 are essential in our analysis. If one needs to obtain data with exactly  the same $\alpha^*$ value as the Gaussian data, one has to use the iterative Schreiber-Schmitz
procedure \cite{SchreiberSchmitz}, where in each iteration the data are (1) Fourier-transformed to $f$ space. Then (2) the Fourier-transformed data are exchanged by the  Fourier-transform of the original Gaussian data and (3) 
 Fourier-transformed back to time space. Finally, (4) these data are exchanged rankwise by the desired non-Gaussian distribution. 
By comparing the simple exchange method with the Schreiber-Schmitz procedure we found that 
both methods yield, for the same global Hurst exponent $\alpha^*\le 1.5$, the same distribution $H(\alpha; \alpha^*,L)$ of local Hurst exponents $\alpha$  as the Gaussian records.

\begin{figure}[]
\begin{center}
\includegraphics[width=8.3cm]{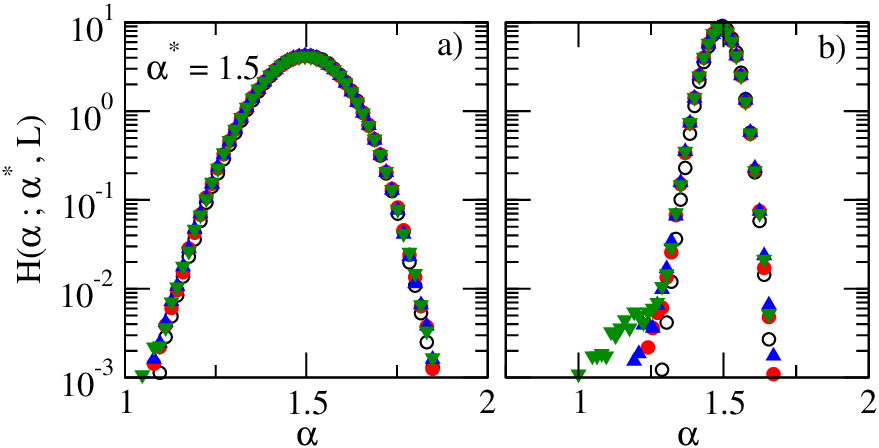}
\caption{(color online) Distribution  $H(\alpha;\alpha^*,L)$ of local Hurst exponents $\alpha$ (obtained by DFA2) in segments of lenght $L=400$ and $L=2200$, for fixed 
global Hurst exponents $\alpha^*=1.5$. $H(\alpha;\alpha^*,L)$ has been obtained for data following a Gaussian distribution (open circles), an exponential distribution (full circles), a symmetric exponential distribution (triangle up), 
and a power law distribution (triangle down). The non-Gaussian data have been generated by the exchange technique.
}
\label{6}
\end{center}
\end{figure}

Figure  6 shows, for $\alpha^*=1.5$ and $L=400$ and 2200,   the distributions  $H(\alpha;\alpha^*,L)$ of local Hurst exponents $\alpha$
 for both Gaussian  and  non-Gaussian data.
Above  $\alpha^*=1.5$,  we were unable to generate long-term correlated records with the considered non-Gaussian distributions. Irrespective of the input value of $\alpha^*$, the output $\alpha^*$ was always close to 1.5, and the distribution of the $\alpha$ values became much broader than for the Gaussian data. 
Accordingly, our analysis of non-Gaussian data is limited to those  $\alpha$ values that are typically absent in long records with $\alpha^*$ above 1.5.
We found that this is the case for subrecords of length $L=400$, 1200, and 2200, for $\alpha$ below  $\alpha_c(L)=1.15$, 1.31, and 1.34, respectively, where the fraction of subrecords originating from  long records with $\alpha^*$ above 1.5, is below $10^{-3}$. Accordingly, our analysis holds only for local Hurst exponents $\alpha$ below $\alpha_c(L)$.
\begin{figure}[]
\begin{center}
\includegraphics[width=8.3cm]{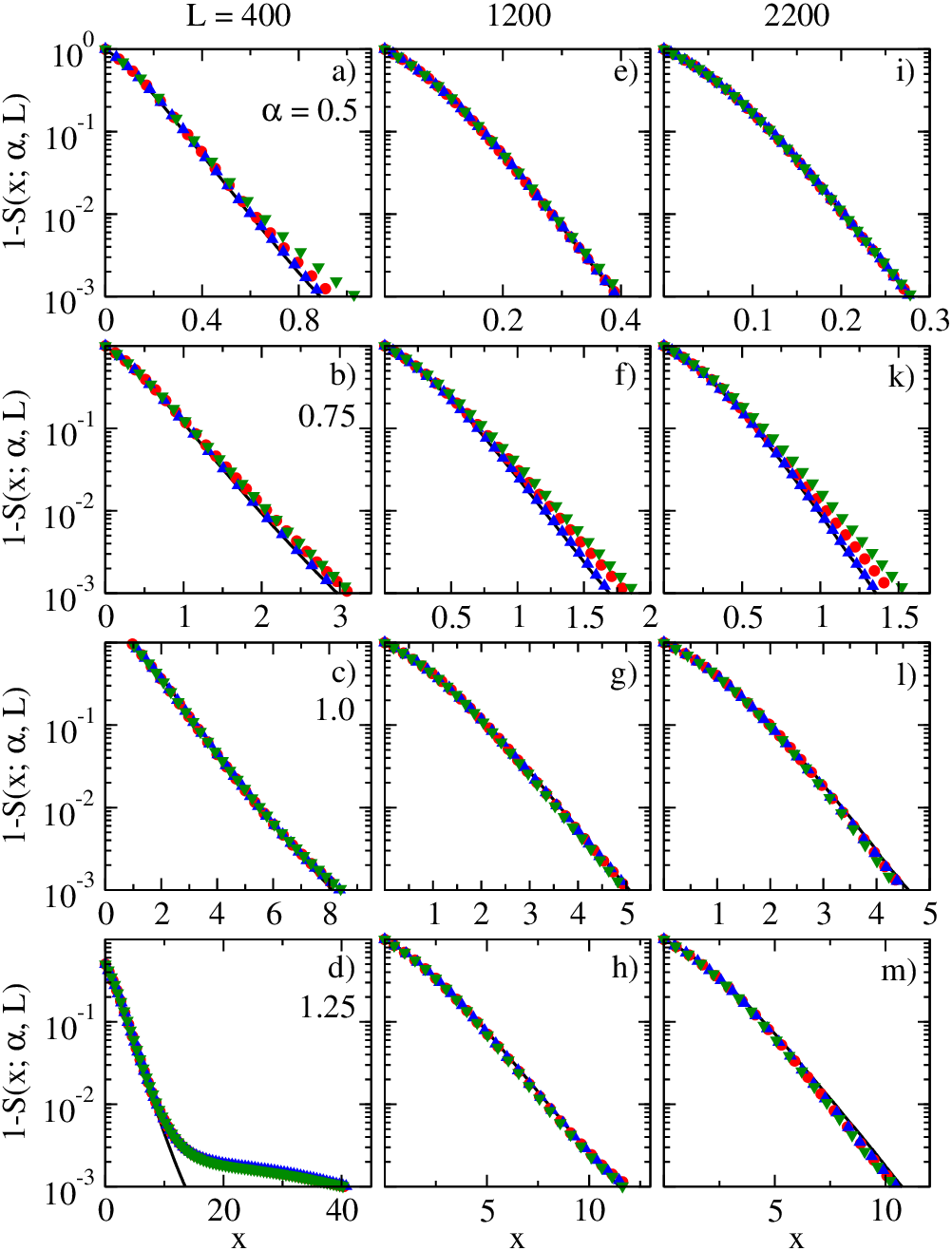}
\caption{(color online) Dependence of the significance $S (x, \alpha,L)$ of the relative trend $x$ on the  distribution of the long-term persistent data. Apart from Gaussian data (continous line, shown also in Fig. 2), we consider data with
 asymmetric exponential distribution (full circle), 
symmetric exponential distribution (triangle up), and power law distribution (triangle down). For convenience, we show $1-S$ as in Fig. 2. The  record lenghts are  $L=400$, 1200, and 2200  (columns $1-3$), and the Hurst exponents are $\alpha=0.5,0.75, 1$, and 1.25 (rows $1-4$). 
}
\label{7}
\end{center}
\end{figure}

Figure 7 shows the significance of  the relative trend $x$ for the 3 non-Gaussian distributions considered here, for $L=400$, 1200, and 2200. The continous line
(difficult to see) is the result for the Gaussian data. The figure shows that for $\alpha$ below 1 (rows $1-3$ in the figure), the data for the two-sided exponential distribution fully coincide
with the Gaussian data, while there are minor deviations for the strongly skewed data. It is important to note that $1-S$ of the Gaussian distribution appears to be a lower bound for $1-S$ of the skewed distributions, this means that the significance of a trend in  long-term correlated Gaussian data represents an upper
bound for the significance. 

Accordingly, the value obtained for $x_{95}$ in Gaussian distributed data represents a lower bound, but the differences between
the different distributions are very small, the largest deviations occur for $\alpha=0.75$ and $L=2200$ where  $x_{95}$  for the skewed power-law distribution
exceeds $x_{95}$ for the Gaussian distribution by less than 5 percent. For $\alpha=1$, $x_{95}$ is the same for all distributions.  For $\alpha=1.25$ (last row)
which is below $\alpha_c$ for $L=1200$ and 2200, the agreement between Gaussian and non-Gaussian data is perfect for $L=1200$ and 2200. For $L=400$,
where $\alpha$ is above $\alpha_c$ and thus should not be considered, Gaussian and non-Gaussian data still collapse for $1-S$ above $10^{-2}$. The shoulder below $10^{-2}$ is an artifact   which originates from records where the original  $\alpha^*$ value was above 1.5.

\section{Example: The Antarctic Byrd Record}
It is straightforward to apply our methodology to observational data. Important applications are climate data (e.g., river flows, precipitation, and temperature data) where one likes to know the significance of trends due to anthropogenic climate change. When considering climate data, it is important to use monthly data where additional short-term dependencies have been averaged out and seasonal trends can be better eliminated than in daily data \cite{Lennartz2011}.

For convenience we consider temperature data. The seasonal trend elimination is done in 2 steps \cite{Kantelhardt2006, Koscielny2006}. In the first step, we substract the monthly seasonal trend to obtain the temperature anomalies ${\tilde y_i, i=1,\dots, L}$. Since the variance of the temperature anomalies may depend on the season,
we divide in the second step the temperature anomalies by the seasonal standard deviation. The resulting dimensionless record ${y_i}$ has unit variance and zero mean. 

Next we perform the  regression analysis  for the $\{y_i\}$ which yields $\Delta$ and $\sigma$ and thus the relative trend $x$. Then we employ DFA2 and obtain the Hurst exponent $\alpha$. From $x$ and $\alpha$ we can estimate the significance $S$ of the temperature trend from Tables I and II as well as the boundary $\pm x_{95}$ of the 95$\%$ significance interval.

Since we divided the temperature anomalies by the seasonal standard deviation to obtain $\{y_i\}$, $\Delta$ and $\sigma$ as well as  the error bars $\pm \Delta_{95}=x_{95}\sigma$ are dimensionless.
To obtain the real trend $\Delta^{\rm real}$ and its real error bars $\pm \Delta_{95}^{real}$ in units of $^oC$, we perform a regression analysis of the temperature anomalies ${\tilde y_i, i=1,\dots, L}$.
To obtain the error bars $\Delta_{95}^{\rm real}$ we use the identity (see \cite{Lennartz2011})
$\Delta_{95}^{\rm real}/\Delta^{\rm real}=\Delta_{95}/\Delta$, 
which then yields
$$\Delta_{95}^{\rm real}=\Delta_{95} \frac{\Delta^{\rm real}}{\Delta}.$$
The resulting minimum and maximum external temperature trends are $\Delta^{\rm real}\pm \Delta_{95}^{\rm real}$.

To show explicitely how our approach  can be used to estimate the significance of a warming trend we consider the monthly (corrected)
Byrd record between 1957 and 2013 that was recently reconstructed by Bromwich et al. \cite{Bromwich2014}.(An earlier version \cite{Bromwich2012} of the Byrd record has been discussed
in \cite{Bunde2014}).
The Byrd station is located in the center of West Antarctica which is one of the fastest warming places on Earth. 
The regression analysis yields $\Delta^{\rm real}=2.02^oC$.
It is obvious that the question of the significance of Antarctic warming is highly relevant, since the warming trend influences the melting of the West Antarctic Ice Shielf and thus contributes to future sea level rise.

Figure 8a shows the fully seasonally detrended Byrd record $y_i$, $i=1, \dots, 684$, where the temperature anomalies have been divided by the seasonal standard deviation. The regression analysis yields $\Delta=0.69$
 and $\sigma=1.04$, yielding $x=0.66$.

Figure 8b shows the result of the DFA2 analysis for $y_i$ (full circles). In the double logarithmic plot, the DFA2 fluctuation function $F(s)$ follows a straight line with exponent $\alpha=0.65$ between $s=10$ and $L/4$. Accordingly, the data are long-term persistent and our methodology applies. 
The exponent $\alpha = 0.65$ is in agreement with earlier estimates \cite{Bunde2014,Bromwich2014}.
We like to note that the temperature anomalies $\tilde y_i$ 
yield to the same Hurst exponent showing explicitely that the second step in the seasonal detrending has no influence on the persistence properties.

\begin{figure}[]
\begin{center} 
\vspace{0.5cm}
\includegraphics[width=8.5cm]{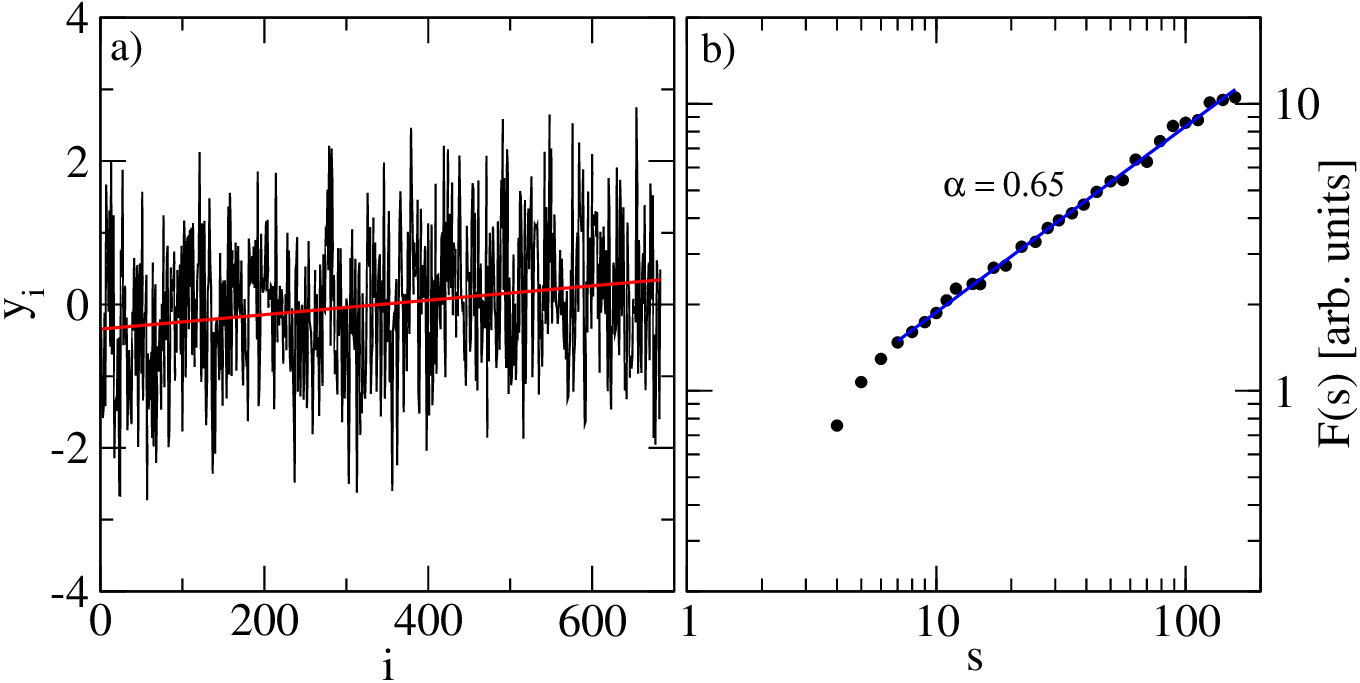}
\caption{(color online) (a) Monthly fully seasonally detrended record $y_i$ at the Byrd station between 1957 and 2013 (black lines) and the corresponding linear regression (red line).
(b) The DFA2 fluctuation function $F(s)$ of the monthly Byrd record $y_i$ (black circles). 
}
\label{8}
\end{center}
\end{figure}

For obtaining the degrees of freedom $l$ and the scaling factor $a$ for $\alpha=0.65$ and $L=684$, we consider the 4th line in Table 1 and 2 and use the respective values for $L=500,600,700$ and 800 for a cubic  interpolation. This gives
$l=9.78$ and $a=0.286$. Inserting these values into Eq. (9) gives  $S=0.953$ and  $x_{95}=0.64$. Accordingly,
the significance of the warming trend at the Byrd station is 95.3\%. The minimum external trend is $0.03^oC$ , while the maximum external trend is $3.99^oC$. 

When Bromwich et al determined these quantities, they used the conventional hypothesis that the {\it annual} linearly detrended temperature data follow an AR(1) process and that $S$ can be obtained from Eqs. (9), (11) and (12) (even though the length of the annual data is too small to make (11) and (12) applicable). For the annual detrended data, they obtained $C(1)=0.075$ and thus $l=49.05$ and $a=0.546$. Inserting the corresponding values into (9) yields $S=0.999$ and $x_{95}=0.997$. The minimum external trend is $0.87^oC$ , while the maximum external trend is $3.09^oC$. 

Accordingly, the significance of the warming trend as well as the minimum external trend has been strongly overestimated by Bromwich et al, while the uncertainty $2x_{95}$ has been underestimated.

\section{Conclusions }  In summary,  we have studied by extensive Monte-Carlo simulations the distribution $P(x;\alpha,L)$ of linear trends in long-term correlated
records of length $L$ that are characterized
by a Hurst exponent $\alpha$ between 0.5 and 1.5
(determined by DFA2). The Hurst exponent was obtained by linear regression from the slope of the regression line  in a double logarithmic representation of the DFA2 fluctuation function $F(s)$ between $s=10$ and $L/4$.  We have considered record lengths $L$ between 400 and 2200, which corresponds, in monthly climate records, to time scales between 33.3 and  183.3 years. In each record we have determined
by linear regression analysis the increase $\Delta$ and
the standard deviation $\sigma$ of the data around the regression line; the ratio $x=\Delta/\sigma$ is the relative trend.

We have extended the earlier analysis \cite{Lennartz2011} in three important directions:

(i) We
found that 
 $P(x;\alpha,L)$ follows, in the whole $x$-range, the student-t distribution with two fit parameters, the scaling parameter $a$ and the effective length $l$. 
This generalizes nicely the known results for white noise and AR(1) processes, where $P$ also follows a student-t distribution, but with different $a$ and $l$ values. For Hurst exponents between 0.5 and 1.1, $l$ depends only on the record length $L$, and not on the Hurst exponent $\alpha$. In the previous work  \cite{Lennartz2011},  the distribution was  approximated by a Gaussian at small $x$ and an exponential at large $x$ values, which  allowed to determine easily the significance of large relative trends.

(ii) In  \cite{Lennartz2011}, only Hurst exponents up to 1.1 could be treated analytically. Here we extend the analytical analysis to $\alpha=1.5$ where the deviations from simple exponential behavior at large $x$ are more pronounced. We also considered slightly smaller and larger record lengths.

(iii) In  \cite{Lennartz2011}, only Gaussian data were considered. Here we have shown explicitely that the results are stable and do hold also for very different, highly skewed distributions.

For applying our methodology to observable data one must be sure that there are no additional short term correlations on short time scales. It is known that in temperature anomalies, there are additional short term correlation on time scales up to 10 days. For river flows, the short term persistence may range up to one month. These short term correlations can be eliminated by averaging the data over short time windows that are larger than the persistence time, i.e. by considering monthly  temperature anomalies and quarter annual river flows. The DFA2 analysis then has to be performed on these averaged records and
the actual data length $L$ is the length of the averaged data set.

In long-term persistent processes it is enough to determine $\alpha$ by DFA2 and use this value to determine $a$ and $l$.
In most previous evaluations of the significance of trends in climate science the significance of trends has been determined from the value of $C(1)$ in annual records, assuming that the significance of trends in an AR(1) record may be a good approximation for the significance of trends in long-term correlated climate records. Our results show that one does not need to rely on this crude approximation (which usually strongly exaggerates the significance) since the estimation of the significance of trends  in long-term correlated records  is not more difficult.

Acknowledgements: We like to thank the Deutsche Forschungsgemeinschaft and the Ministry of Education and Science of the Russian Federation for financial support.

\section{Appendix}
We have shown in this article that the probability $P(x;\alpha,L)$ that in a long-term persistent record of length $L$, characterized by the Hurst exponent $\alpha$, a relative trend of strength $x$ occurs, has the form of a student-t distribution,
 $$P(x;\alpha,L)= \frac{\Gamma(\frac{l+1}{2})}{\Gamma(\frac{l}{2})\sqrt{\pi l}a} \left(1+\frac{(x/a)^2}{l}\right)^{-\frac{l+1}{2}},$$
with the effective length
 $l$
 and the scaling parameter $a$. The related trend significance is  $$S(x;\alpha,L)= 2\frac{x}{a}\frac{\Gamma(\frac{1}{2}(l+1)}{\sqrt{\pi l}\Gamma(\frac{l}{2})} 
 {}_2 F_1 \left(\frac{1}{2},\frac{1}{2}(l+1);\frac{3}{2};-\frac{(x/a)^2}{l}\right).$$ Tables I and II list the  effective lengths $l$ and the scaling factor $a$ as function of the DFA2 Hurst exponent $\alpha$ and the record length $l$. 

\begin{table*}
\begin{tabular}{ | c |c| c| c| c| c| c| c| c| c |c |c |c |c |}
\hline 
\hline 
\rule[2.1mm]{0mm}{2.1mm} 
$\alpha  \backslash L$ & L400	&L500	 &  L600  &   L700 &   L800  &  L900 &   L1000&   L1200 &  L1400 &  L1600 &  L1800 &   L2000 &   L2200  \\[0.5ex]
\hline
0.50 &    7.60  &   8.50 &    9.24 &	9.87 &    10.41  &  10.88  &  11.31 &	12.05 &   12.67 &   13.21 &   13.69 &	14.11 &   14.50\\
0.55 &    7.60  &   8.50 &    9.24 &	9.87 &    10.41  &  10.88  &  11.31 &	12.05 &   12.67 &   13.21 &   13.69 &	14.11 &   14.50\\
0.60 &    7.60  &   8.50 &    9.24 &	9.87 &    10.41  &  10.88  &  11.31 &	12.05 &   12.67 &   13.21 &   13.69 &	14.11 &   14.50\\
0.65 &    7.60  &   8.50 &    9.24 &	9.87 &    10.41  &  10.88  &  11.31 &	12.05 &   12.67 &   13.21 &   13.69 &	14.11 &   14.50\\
\hline
0.70 &    7.60  &   8.50 &    9.24 &	9.87 &    10.41  &  10.88  &  11.31 &	12.05 &   12.67 &   13.21 &   13.69 &	14.11 &   14.50\\
0.75 &    7.60  &   8.50 &    9.24 &	9.87 &    10.41  &  10.88  &  11.31 &	12.05 &   12.67 &   13.21 &   13.69 &	14.11 &   14.50\\
0.80 &    7.60  &   8.50 &    9.24 &	9.87 &    10.41  &  10.88  &  11.31 &	12.05 &   12.67 &   13.21 &   13.69 &	14.11 &   14.50\\
0.85 &    7.60  &   8.50 &    9.24 &	9.87 &    10.41  &  10.88  &  11.31 &	12.05 &   12.67 &   13.21 &   13.69 &	14.11 &   14.50\\
\hline
0.90 &    7.60  &   8.50 &    9.24 &	9.87 &    10.41  &  10.88  &  11.31 &	12.05 &   12.67 &   13.21 &   13.69 &	14.11 &   14.50\\
0.95 &    7.60  &   8.50 &    9.24 &	9.87 &    10.41  &  10.88  &  11.31 &	12.05 &   12.67 &   13.21 &   13.69 &	14.11 &   14.50\\
1.00 &    7.60  &   8.50 &    9.24 &	9.87 &    10.41  &  10.88  &  11.31 &	12.05 &   12.67 &   13.21 &   13.69 &	14.11 &   14.50\\
1.05 &    7.60  &   8.50 &    9.24 &	9.87 &    10.41  &  10.88  &  11.31 &	12.05 &   12.67 &   13.21 &   13.69 &	14.11 &   14.50\\
\hline
1.10 &    7.60  &   8.50 &    9.24 &	9.87 &    10.41  &  10.88  &  11.31 &	12.05 &   12.67 &   13.21 &   13.69 &	14.11 &   14.50\\
1.15 &    7.06  &   7.69 &    8.86 &	9.32 &    9.63   &  10.51  &  11.08 &	11.76 &   13.12 &   12.50 &   13.07 &	14.24 &   14.44\\
1.20 &    6.95  &   7.50 &    8.27 &	9.54 &    9.21   &  9.63   &  10.28 &	11.14 &   11.91 &   11.92 &   12.48 &	13.13 &   14.26\\
1.25 &    6.41  &   7.32 &    7.91 &	8.74 &    8.61   &  9.27   &  9.85  &	10.55 &   11.21 &   11.34 &   11.59 &	12.44 &   12.94\\
\hline
1.30 &    6.25  &   6.68 &    7.60 &	8.11 &    8.28   &  8.55   &  9.28  &	9.90  &   10.21 &   10.36 &   10.74 &	11.26 &   11.03\\
1.35 &    5.93  &   6.31 &    7.29 &	7.68 &    7.83   &  8.21   &  8.74  &	9.12  &   9.88  &   9.88  &   10.26 &	10.40 &   10.40\\
1.40 &    5.47  &   5.80 &    6.63 &	7.21 &    7.44   &  7.55   &  7.93  &	8.49  &   8.94  &   8.78  &   9.02  &	9.48  &   9.74\\
1.45 &    5.19  &   5.54 &    6.33 &	6.83 &    6.83   &  7.14   &  7.58  &	7.82  &   8.62  &   8.18  &   8.43  &	8.80  &   9.01\\
1.50 &    4.77  &   5.10 &    5.88 &	6.45 &    6.33   &  6.76   &  6.88  &	7.42  &   7.48  &   7.55  &   7.76  &	8.25  &   8.41\\
\hline
\end{tabular}

\caption{Effective length $l(\alpha, L)$ in  in the  trend  distribution $P$
and the trend significance $S$ in long term correlated data with Hurst exponent $\alpha$ and record length $L$.}

\end{table*}

\begin{table*}[ht]
\vspace{0.5cm}
\renewcommand\arraystretch{1.03}

\centering

\begin{tabular}{ | c |c| c| c| c| c| c| c| c| c |c |c |c |c |}
\hline 
\hline 
\rule[2.1mm]{0mm}{2.1mm} 
$\alpha \backslash L$ & L400	&L500	 &  L600  &   L700 &   L800  &  L900 &   L1000&   L1200 &  L1400 &  L1600 &  L1800 &   L2000 &   L2200  \\[0.5ex]
\hline
0.50   &  0.174 &  0.162 &  0.133 &  0.124 &  0.115  & 0.107 &  0.104 &  0.092  & 0.086  & 0.081  & 0.076  & 0.072   & 0.068 \\
0.55   &  0.227 &  0.212 &  0.177 &  0.165 &  0.154  & 0.145 &  0.140 &  0.126  & 0.117  & 0.111  & 0.105  & 0.100   & 0.096 \\
0.60   &  0.292 &  0.275 &  0.232 &  0.218 &  0.205  & 0.193 &  0.188 &  0.171  & 0.160  & 0.153  & 0.145  & 0.139   & 0.133\\
0.65   &  0.371 &  0.352 &  0.300 &  0.284 &  0.268  & 0.256 &  0.249 &  0.227  & 0.215  & 0.206  & 0.196  & 0.190   & 0.182\\
\hline
0.70   &   0.465&   0.447&   0.385&   0.368&   0.348 &  0.332&   0.325&   0.301 &  0.286 &  0.275 &  0.265 &  0.256  & 0.247\\
0.75   &   0.577&   0.556&   0.486&   0.466&   0.446 &  0.428&   0.420&   0.392 &  0.376 &  0.364 &  0.351 &  0.341  & 0.332\\
0.80   &   0.706&   0.687&   0.606&   0.585&   0.564 &  0.543&   0.537&   0.504 &  0.486 &  0.473 &  0.460 &  0.448  & 0.438\\
0.85   &   0.854&   0.835&   0.745&   0.726&   0.703 &  0.681&   0.675&   0.639 &  0.621 &  0.609 &  0.593 &  0.582  & 0.568\\
\hline
0.90   &   1.021&   1.006&   0.906&   0.883&   0.864 &  0.841&   0.833&   0.800 &  0.778 &  0.769 &  0.752 &  0.742  & 0.726\\
0.95   &   1.208&   1.198&   1.088&   1.069&   1.051 &  1.026&   1.021&   0.981 &  0.961 &  0.952 &  0.937 &  0.924  & 0.917\\
1.00   &   1.417&   1.410&   1.291&   1.276&   1.256 &  1.233&   1.229&   1.187 &  1.171 &  1.167 &  1.151 &  1.142  & 1.100\\
1.05   &   1.646&   1.648&   1.517&   1.506&   1.488 &  1.466&   1.456&   1.420 &  1.409 &  1.404 &  1.390 &  1.340  & 1.366\\
\hline
1.10   &   1.900&   1.915&   1.776&   1.758&   1.745 &  1.722&   1.716&   1.681 &  1.666 &  1.670 &  1.657 &  1.590  & 1.633\\
1.15   &   2.138&   2.133&   2.024&   2.016&   1.992 &  1.985&   1.993&   1.953 &  1.971 &  1.938 &  1.937 &  1.938  & 1.925\\
1.20   &   2.427&   2.420&   2.288&   2.327&   2.273 &  2.261&   2.268&   2.248 &  2.244 &  2.238 &  2.237 &  2.231  & 2.238\\
1.25   &   2.695&   2.758&   2.584&   2.615&   2.568 &  2.569&   2.580&   2.556 &  2.559 &  2.561 &  2.548 &  2.552  & 2.557\\
\hline
1.30   &   3.045&   3.044&   2.917&   2.927&   2.906 &  2.884&   2.915&   2.893 &  2.883 &  2.887 &  2.886 &  2.887  & 2.857\\
1.35   &   3.404&   3.411&   3.286&   3.287&   3.264 &  3.252&   3.279&   3.242 &  3.267 &  3.261 &  3.263 &  3.239  & 3.228\\
1.40   &   3.758&   3.750&   3.621&   3.662&   3.665 &  3.619&   3.637&   3.640 &  3.643 &  3.626 &  3.630 &  3.624  & 3.635\\
1.45   &   4.202&   4.234&   4.052&   4.102&   4.062 &  4.060&   4.101&   4.042 &  4.097 &  4.061 &  4.055 &  4.065  & 4.060\\
1.50   &  4.666 &   4.679 &  4.538 &  4.615 &  4.534  & 4.554 &  4.546 &  4.550  & 4.514  & 4.535  & 4.530  & 4.560   &4.561\\
\hline
\end{tabular}
\caption{Scaling factor $a(\alpha, L)$ in the  trend  distribution $P$
and the trend significance $S$ in long term correlated data with Hurst exponent $\alpha$ and record length $L$.}
\label{table2}
\end{table*}
\section*{ }


\end{document}